%% file: main.tex
\documentclass[runningheads]{src/llncs}
\input{src/packages}

\input{src/macros}

\begin{document}
\title{TrustOps: Continuously Building Trustworthy Software}

\titlerunning{TrustOps}

\hide{
\author{
Eduardo Brito\inst{1}\orcidID{0009-0002-9996-6333} \and
Fernando Castillo\inst{2}\orcidID{0009-0003-6835-8711} \and
Pille Pullonen-Raudvere\inst{1}\orcidID{0000-0002-3255-7001} \and 
Sebastian Werner\inst{2}\orcidID{0000-0001-8051-7226}
}

\authorrunning{Eduardo Brito et al.}

\institute{Cybernetica AS, Estonia 
M\"aealuse 2/1, 12618 Tallinn, Estonia\\
\email{\{eduardo.brito, pille.pullonen-raudvere\}@cyber.ee} \and
Information Systems Engineering, TU Berlin, Einsteinufer 17, Germany\\
\email{\{fc,sw\}@ise.tu-berlin.de}
}

}

\maketitle

\begin{abstract}

\input{sections/00_abstract}

\end{abstract}

\newcommand{\cop}[1]{%
    \texttt{\detokenize{#1}}
}

\section{Introduction}\label{ch:intro}

\input{sections/01_intro}

\section{Related Work}\label{ch:rw}
\input{sections/02_rw}

\section{Evidence-based Trustworthiness}\label{ch:main}
\input{sections/03_objectives_evidence}

\section{Application Scenarios}\label{ch:senarios}
\input{sections/04_scenarios}

\section{TrustOps}\label{ch:phases}

\input{sections/05_trustops_phases}

\section{Research and TrustOps Adoption Challenges}\label{ch:challenges}
\input{sections/06_research_challenges}

\section{Conclusion}\label{ch:fin}
\input{sections/07_conclusion}

\section*{Acknowledgements}{Funded by the European Union (TEADAL, 101070186). Views and opinions expressed are, however, those of the author(s) only and do not necessarily reflect those of the European Union. Neither the European Union nor the granting authority can be held responsible for them.}

\bibliographystyle{src/splncs04}
%using minfied version
\bibliography{refs}

\end{document}

%% file: src/packages.tex
\usepackage{cite}
\usepackage{amsmath,amssymb,amsfonts}
\usepackage{algorithmic}
\usepackage{graphicx}
\usepackage{textcomp}
\usepackage{xcolor}
\usepackage{csquotes}
\usepackage{atbegshi}
\usepackage{hyperref}

\usepackage{cleveref}
\usepackage{csquotes}

%% file: src/macros.tex
\newcommand{\hide}[1]{#1}
\newcommand{\evidence}{evidence}

%% file: sections/00_abstract.tex
Software services play a crucial role in daily life, with automated actions determining access to resources and information. 
Trusting service providers to perform these actions fairly and accurately is essential, yet challenging for users to verify. 
Even with publicly available codebases, the rapid pace of development and the complexity of modern deployments hinder the understanding and evaluation of service actions, including for experts.
Hence, current trust models rely heavily on the assumption that service providers follow best practices and adhere to laws and regulations, which is increasingly impractical and risky, leading to undetected flaws and data leaks.

In this paper, we argue that gathering verifiable evidence during software development and operations is needed for creating a new trust model.
Therefore, we present TrustOps, an approach for continuously collecting verifiable evidence in all phases of the software life cycle, relying on and combining already existing tools and trust-enhancing technologies to do so. 
For this, we introduce the adaptable core principles of TrustOps and provide a roadmap for future research and development.

%% file: sections/00_keywords.tex
trustworthy software, continuous software engineering, \\evidence-based life cycle

%% file: sections/01_intro.tex
Software services pervade many areas of daily life where actions, sometimes automatically, decide how we can access resources, e.g., payment services decide if we are creditworthy enough to buy train tickets, or social media websites decide which news to show us. 
Users must trust that service providers act fairly and that actions are performed correctly. However, users hardly have the means to evaluate how these services were built or provided to them~\cite{huhnlein2022futuretrust}. 
Even if publicly available codebases were present, a typical user cannot evaluate them. 
Magnified by the accelerated pace of modern development practices like DevOps\cite{bass2015devops}, the complexity of these distributed systems, consisting of multiple, independently developed and configured components, makes independent verification of service delivery even challenging for experts.

Hence, current trust models largely rely on the assumption that service providers follow best practices, adhere to laws and regulations, and that public audit processes will catch issues before they impact users. 
This reliance on a few large organizations for trust is becoming increasingly impractical and risky, introducing the potential for undetected exploited flaws, unintentional data leaks, or risks due to supply chain attacks, such as the recent XZ vulnerability\footnote{(CVE-2024-3094) \url{https://nvd.nist.gov/vuln/detail/CVE-2024-3094}}.
Not only must developers employ practices like security tests and well-defined processes, but users should also be able to verify that a service they use was tested and followed all established procedures.

We argue that these challenges necessitate a paradigm shift towards verifiable development and operational processes, where stakeholders can independently verify the integrity and authenticity of such actions. 
This new approach, which we term TrustOps, aims to address the gaps left by current methodologies, such as DevSecOps~\cite{sanchez2020security} and Continuous Compliance\cite{fitzgerald2017continuous}. 
While these approaches incorporate security practices into the development lifecycle, they do not fully address the need for comprehensive, publicly verifiable evidence of the entire service development process. 
Therefore, we see TrustOps as an addition to these methodologies, focused on ensuring that these practices cannot be bypassed when delivering services. 
Moreover, we do not attempt to, or assume that risks or flaws can be entirely prevented, but the chain of events that caused a flaw can be followed and discovered early.

TrustOps advocates for implementing or using existing robust, auditable mechanisms that track changes, combine and aggregate development evidence (e.g., change logs, reviews), and collect and provide evidence of test executions and policy actions. 
The evidence, collected at each stage of the development process, must be independently and automatically verifiable to create an easy-to-consume audit trail. 
Particularly, in this paper, we:
\begin{itemize}
    \item Present evidence-based trustworthiness principles that enable the build-up of authentic, attestable, and actionable evidence during the software life cycle.
    \item Detail the use and application of TrustOps, based on examples and scenarios, along typical DevOps phases.
    \item Highlight a set of research objectives needed to foster, improve, and adopt TrustOps.
\end{itemize}

In the remainder of this paper we first present related work in \cref{ch:rw}. Then, in \cref{ch:main}, we introduce the main processes of evidence-based trustworthiness, before providing exemplary application scenarios in \cref{ch:senarios}. Lastly, in \cref{ch:phases}, the evidence life cycle is mapped to DevOps, defining the TrustOps approach, with research challenges outlined in \cref{ch:challenges}, before concluding in \cref{ch:fin}.

%% file: sections/02_rw.tex
Trust, as defined by various scholars \cite{huhnlein2022futuretrust, mayer1995integrative, hou2023systematic}, is a complex interplay of beliefs, expectations, and reliability. In software systems, trust is the reliance on software's ability to meet specified requirements, even amid uncertainty. It is grounded in the expectation of consistent behavior, despite unpredictable conditions \cite{huhnlein2022futuretrust, hou2023systematic}. Software trustworthiness involves building confidence in its ability to fulfill intended functions, ensuring reliability, security, and consistency over time \cite{heiss2023trustworthy, hou2023systematic}. The multifaceted nature of trust in software reflects sociological, psychological, philosophical, and computational perspectives \cite{huhnlein2022futuretrust}, making it a socio-technical phenomenon shaped by human perceptions and organizational structures \cite{ting2021trust}.

Adherence to industry standards has been contributing to enhance perceived trust and establish globally recognized quality assurance~\cite{hou2023systematic, fitzgerald2017continuous}. ISO, IEC, IEEE standards serve as benchmarks for best practices in information security and privacy, aiming, for instance, at protecting Personally Identifiable Information (PII), deploying Privacy Information Management Systems (PIMS), integrating trustworthy elements into software, establishing security for industrial automation and control systems, or standardizing life cycle processes like DevOps and Agile development~\cite{ISO/IEC27018:2019,ISO22376:2023,IEC62443,ISO/IEC/IEEE32675:2022}.

Additionally, trust mechanisms integrated into mainstream development processes like DevOps have been gaining increasing attention. Various approaches, including DevSecOps, DevPrivOps, and VeriDevOps, have emerged to embed robust security measures, privacy considerations, and trust-building protocols throughout the software life cycle. DevSecOps emphasizes security testing integration at different stages \cite{sanchez2020security}, DevPrivOps incorporates privacy engineering \cite{grunewald2021cloud}, while VeriDevOps enhances automation for system protection \cite{enoiu2023veridevops}. Complemented by EU regulations for electronic services, adherence to standards, development methodologies, and compliance protocols are known to enhance trustworthiness, mitigate risks, and build robust stakeholder relationships across the software supply chain \cite{hou2023systematic, fitzgerald2017continuous}. These practices will be explored further in this paper, with TrustOps aiming to unify and streamline trust automation, integrated with generalized methodologies like DevOps, across the entire software life cycle.

%% file: sections/03_objectives_evidence.tex
In this section, we first outline the objectives we seek in establishing trustworthy principles in the software life cycle, and then describe the evidence-based process that builds the foundations of TrustOps. 

\subsection{Objectives}

\begin{figure}[h]
    \centering

    \includegraphics[width=0.7\textwidth]{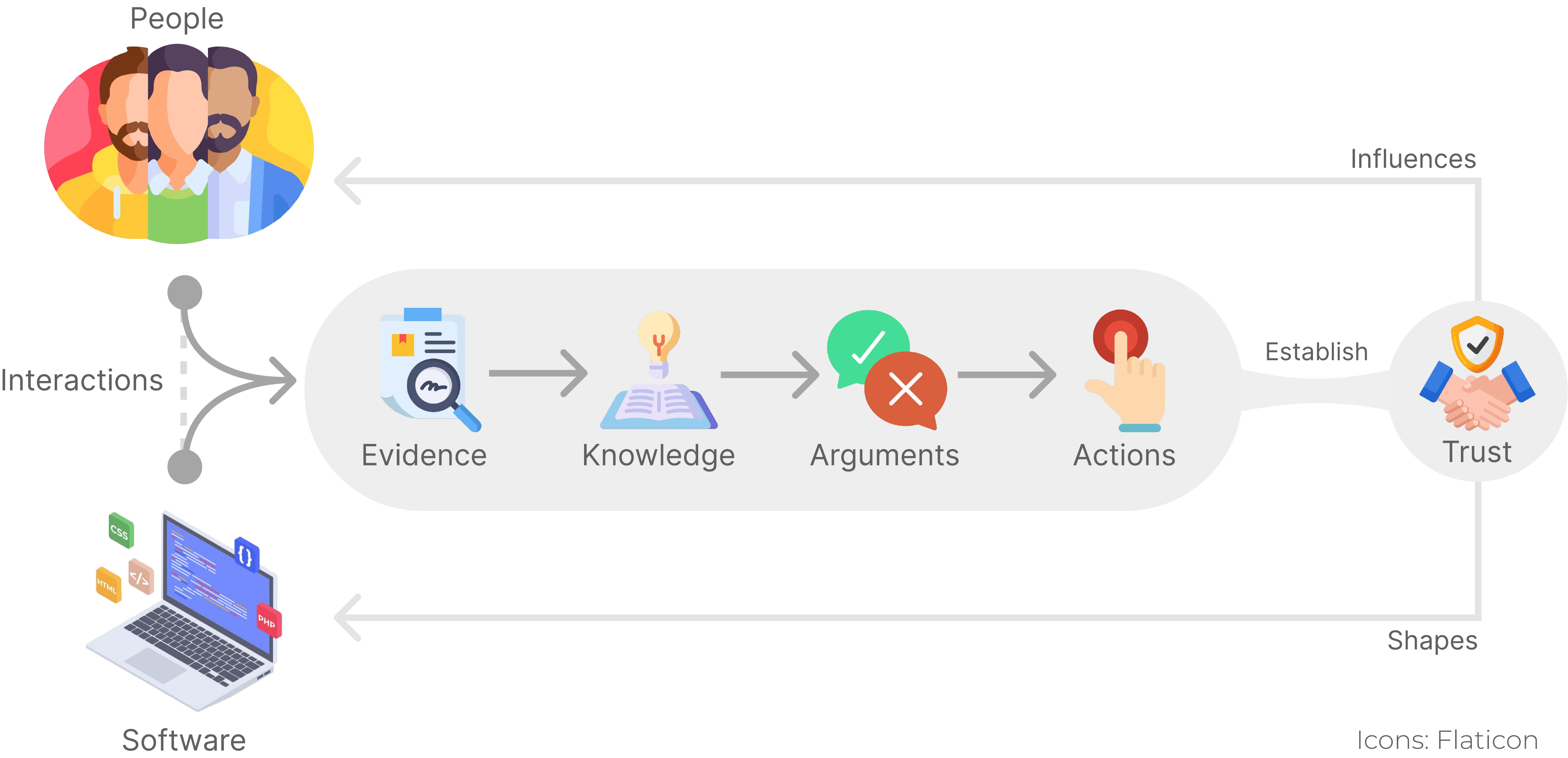}
    \caption{Trust as a socio-technical phenomenon~\cite{ting2021trust}, result of an evidence-based process, that influences people, software, and their interactions~\cite{nistsp800160v1r1, fitzgerald2017continuous}.}
    \label{fig:trust-phenomenon}
\end{figure}

The core objective of TrustOps is to create verifiable development practices, thus enabling all stakeholders in the development and usage process to verify who, when, where, why, or how decisions are made, how changes are rolled out to users, and how users consume these changes. Ron Ross et al.~\cite{nistsp800160v1r1} emphasize the importance of evidence in building trustworthiness.
Evidence is, therefore, a fundamental pillar supporting trust and pertains to the availability of verifiable and credible information that vindicates software's claims and behavior \cite{nistsp800160v1r1, huhnlein2022futuretrust}.
This assurance is further fortified by the roles played by authentication and integrity in substantiating the evidence \cite{heiss2023trustworthy, hou2023systematic}. 
Moreover, while verifiability facilitates systematic scrutiny of trust claims \cite{delignat2023should}, authorization governs access rights \cite{ manuel2015trust}, and distribution, as articulated by \cite{dauterman2022reflections}, enhances security and reliability guarantees.

Interactions between people and software often exhibit an interplay of these concepts, as illustrated in \Cref{fig:trust-phenomenon}. This sets possible avenues for the automated trust paradigm that TrustOps aims to establish. Such paradigm is finding its basis in the development of robust electronic and cryptographic mechanisms with ability to establish secure and verifiable interactions \cite{delignat2023should, dauterman2022reflections}. Through cryptographic protocols, different trust models may be instantiated, allowing automated verification of software claims, while ensuring that the truth is not easily manipulated or fabricated. The principles outlined in \cite{dauterman2022reflections}, such as the use of secure hardware, consensus mechanisms, or append-only logs, lay the groundwork for constructing distributed-trust systems. As technology advances, other automated trust mechanisms may further enhance the efficiency, security, and reliability of trustworthy software, reducing the reliance on subjective human assessments and further solidifying trust in the digital landscape \cite{delignat2023should}. TrustOps also aims to provide verifiability without compromising confidentiality and privacy. 
With recent improvements in practical cryptography~\cite{delignat2023should, eberhardt2021scalable}, such as Trusted Execution Environments (TEEs) or Zero-Knowledge Proofs (ZKPs), these so far opposing objectives can be met. TEEs use hardware-based secure computing to process data securely, while ZKPs allow proving the truth of a statement without revealing the actual data. In TrustOps, we highlight that finding a balance between \textbf{auditability}, \textbf{verifiability} and \textbf{confidentiality} is key. Moreover, not every software system requires the full range or full depth of these qualities. Thus, TrustOps should also be flexible in the range of qualities it can add to software development and operation.

\subsection{The Life Cycle of Evidence}

\begin{figure}
    \centering
    \includegraphics[width=0.7\textwidth]{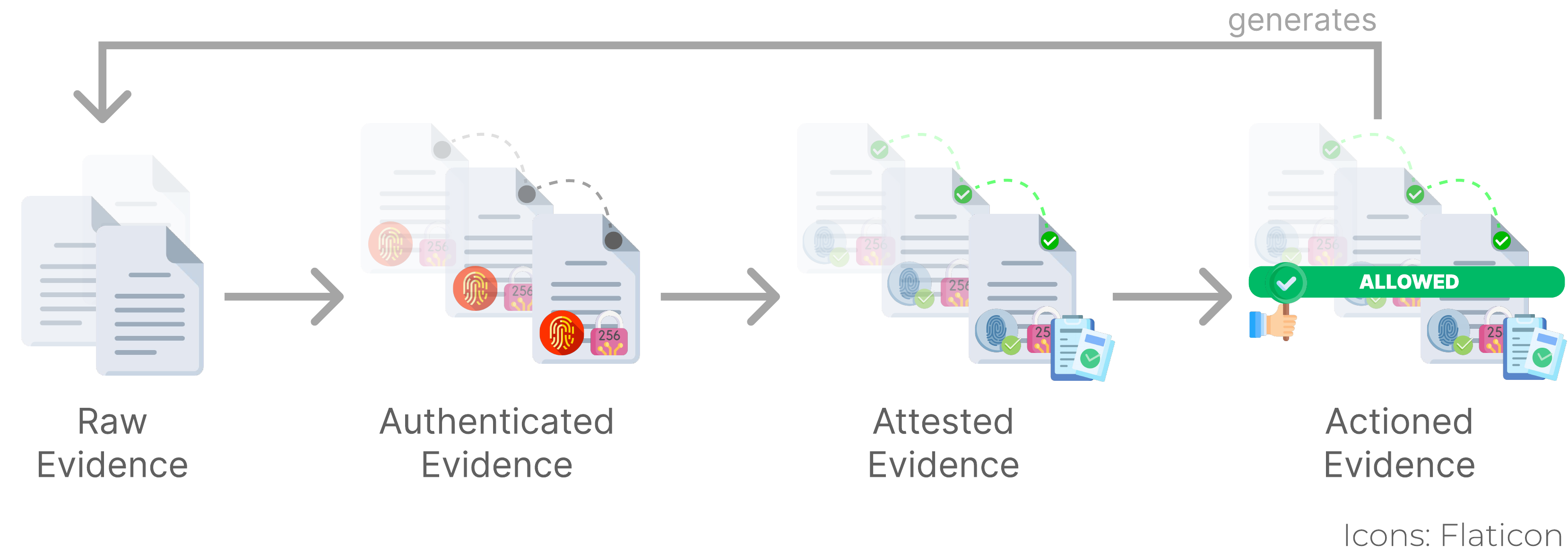}
    \caption{The continuous life cycle process of enhancing evidence. Interactions with software generate evidence that can be authenticated to produce meaningful knowledge and attest to claims about the software's development or operation, leading to informed actions that can, in turn, produce more evidence.}
    \label{fig:lifecycle-evidence}
\end{figure}

TrustOps stands on an evidence-based approach that depends on the evolution and enhancement of evidence, as depicted in \Cref{fig:lifecycle-evidence}, for continuously building trustworthiness, throughout the software life cycle.

\paragraph{\textbf{From Raw to Authenticated Evidence.}}

At the core of TrustOps is the principle of evidence generation, emphasizing the generation of meaningful data and artifacts, throughout the software life cycle. Evidence refers to any information attesting to software events or behavior \cite{nistsp800160v1r1}. For instance, some cases may benefit from tracking code changes and commit metadata, others from storing build logs, deployment details, infrastructure configurations, monitoring data, or end-user feedback. The goal is to gather comprehensive and diverse sets of records, promoting transparency and accountability, to meet eventual auditability and verifiability demands.

Throughout its generation, raw evidence should be enhanced with properties that produce knowledge, such as metadata, or attributes indicating who, when, or where specific events happened. However, a generally trustworthy system should ensure that the generated evidence is not only extensive and meaningful but also genuine. Current approaches establish that such trust should be anchored and validated through identity verification, maintaining the authenticity of critical evidence and processes~\cite{hou2023systematic}. Examples, taken from DevSecOps, include authenticating commits and having branch rules verifying contributor identities in version control systems~\cite{myrbakken2017devsecops}.

Integrity, another principle linked to trust, refers to the unicity, consistency, and reliability of data and processes \cite{heiss2023trustworthy}. It ensures that data is unique to any moment and remains unaltered over time, and that related processes intentionally operate with dependability over past knowledge. Accurate, reliable, and tamper-proof information generation is crucial for informed decisions and structural risk mitigation \cite{nistsp800160v1r1}. Both authentication and integrity may rely on cryptographic mechanisms to create unique digital footprints, ensure data consistency, and prevent or, at least, make evident any unauthorized alterations \cite{dauterman2022reflections}. Digital signatures, proofs of properties, and other mechanisms can be applied to source code, configuration files, documentation, build artifacts, telemetry, and more.

Consequently, this identity-centric approach enables authorization, accountability, and secure evidence distribution, facilitating process automation and the trustworthy generation of new knowledge about software systems \cite{ting2021trust}. In short, evidence not only forms extensive knowledge, but, through these authentication and integrity mechanisms, can also become genuine and consistently reliable, turning into authenticated evidence that can then be attested.

\paragraph{\textbf{From Authenticated to Attested Evidence.}}

Following its enhancement, authenticated evidence becomes a valuable asset for informed decision-making, supporting the development process. Authentic evidence can be used to make claims about software properties or events, which can be verified by interested parties, like users or developers, to ensure specific claims are satisfied or establish agreements and coordination over operations~\cite{nistsp800160v1r1}. Nevertheless, storing evidence must balance trust demands, privacy considerations, or operational capacity, making evidence collection, enhancement, and management central to TrustOps.

Verifiability refers to stakeholders' ability to independently and succinctly validate evidence, actions, and results, ensuring credibility and trustworthiness through transparent means \cite{delignat2023should}. Verifiability follows a proof paradigm, where one side convinces the other of the veracity of properties of data or software. Trust begins with authenticating evidence, followed by automated and attestable proofs using technologies like TEEs or ZKPs, to enable, for instance, automated audits, external artifact validation, or reproducible compliance checks~\cite{fitzgerald2017continuous}. Enabling continuous verifiability allows for automated argumentation and attestation over the trustworthiness of software, enhancing the system's security, reliability, and scalability, and impacting both its objective and perceived trust \cite{eberhardt2021scalable}.

Consensus is also crucial for trustworthiness, especially in distributed settings with different stakeholders holding distributed information, resources and responsibilities~\cite{dauterman2022reflections}. Consensus involves stakeholders collectively agreeing on the validity of arguments over software functionality or behavior. The goal is to ensure software systems meet desired trust levels for all parties, by attesting to shared evidence and agreeing on the validity of claims and actions \cite{heiss2023trustworthy, nistsp800160v1r1}. Examples are database replication, distributed ledgers, decentralized key management, or version control systems, which store and act upon shared evidence, enabling its distributed attestation.

\paragraph{\textbf{From Attested to Actioned Evidence.}}

Once evidence is attested and consensus is reached in distributed settings, informed and automated actions can follow. A typical subsequent action, familiar to the software landscape, is authorization, which involves defining and granting permissions to verified entities, ensuring that only approved actors or processes access specific resources \cite{manuel2015trust}. Code repositories, deployment configurations, or monitoring infrastructure are resources that usually require policy definitions and access control mechanisms. Authorization entails granting permissions based on attested evidence, conceptualizing roles, policies, or attributes \cite{nistsp800160v1r1}, and access control mechanisms ensure these permissions are upheld and not circumvented. Following the integration of authenticated and attested evidence, one can, for instance, define the set of actions and access policies for execution environments or roll-out permissions, like ensuring only known, tested, or verified versions of software can be deployed to production. In addition to authorization, further possible actions can be taken upon attested evidence. For accountability, observability, or explainability purposes, attested evidence can also build a chain of verifiable actions that led to or created an observed outcome. Ultimately, such actions feed the continuous cycle by generating more evidence as their output.\\[0.2em]

To conclude, evolving and enhancing evidence shapes trust, security, and reliability throughout the software life cycle. The outlined evidence-based processes address current trustworthiness challenges and lay the groundwork for adapting to future complexities. The next sections show how these processes can be applied in various scenarios, extending methodologies like DevOps to materialize TrustOps as a continuous set of processes for building trustworthy software.

%% file: sections/04_scenarios.tex
In this section, we introduce three scenarios that showcase motivating examples and different possibilities of integrating the evidence-based concepts of TrustOps throughout the life cycle of various types of software.

\subsection{Trustworthy Open Software}

Open-source tools, libraries, and web repositories inherently promote transparency due to their open nature. Already aligned with a share of TrustOps vision — the source code, collaboration data, contributors and their identities are generally public, roles and rules are publicly assigned, and certain surface metrics are openly scrutinizable. Genuine and trustworthy evidence is, therefore, a widely available byproduct of the public and open collaboration, grounded on the authenticity, integrity, and distributed guarantees provided by modern version control systems. However, some repositories, wielding considerable influence over critical internet systems, deviate from best practices in maintainability and security. Successful projects like the Linux kernel or OpenSSL serve as a positive example, showing transparency and security commitment despite challenges like the Heartbleed vulnerability\footnote{\url{https://heartbleed.com/}}. Conversely, the EventStream incident in the Node.js ecosystem~\footnote{\url{https://blog.npmjs.org/post/180565383195/details-about-the-event-stream-incident}}, or the Log4Shell vulnerability in Java\footnote{\url{https://cve.mitre.org/cgi-bin/cvename.cgi?name=CVE-2021-44228}}, highlight the need for stringent measures and community vigilance. 

To ease this, maintainers may be required to follow DevSecOps to integrate security practices into every stage of the CI/CD pipeline, promoting early detection and remediation of security issues, continuous monitoring, and compliance with security standards~\cite{myrbakken2017devsecops}. Yet, users have to actually review this open activity to truly benefit from all this public evidence. Hence, a more comprehensive and consequent application of TrustOps may reduce or automate the needed review activity. Further trust enhancements could be integrated, for instance, by pipelining these CI/CD processes within TEEs, with public attestation of the testing and release artifacts, and assurances that all security controls, compliance checks, and vulnerability assessments were properly executed.
These trust enhancements and their effects may eventually propagate across the entire software supply chain, as elaborated next. Nevertheless, consideration should be given to finding a balance between heightened trust measures and preserving the collaborative, open, fast-evolving nature of these projects.

\subsection{Enhancing Trust in Service Ecosystems}

In today's service ecosystems, users can select from a variety of hosted open-sourced services, allowing them to review the code and benefit from public scrutiny unparalleled with close-sourced options. Here, TrustOps can play a critical role in extending this increased trust also into the operation of these open-sourced services.
Emerging regulation often necessitate demonstrating adherence to specific requirements, such as the storage and processing of all data of users in the EU within European servers~\cite{gdpr2016a}. Applying TrustOps in the operation phase can automate the attestation of such requirements, for example, by collecting and providing attested evidence from the deployment environment for services like MongoDB Atlas, an open-source database service offered by MongoDB.com and other cloud vendors such as Google\footnote{\url{https://cloud.google.com/mongodb}}.

Today, each service vendor delineates Service Level Agreements (SLAs) and contractual terms governing data treatment, processors, and hosting locations. However, users inherently rely on trust in these vendors, which may be hard to attest. While the utilization of open-source software may instill confidence, the veracity of vendor implementation remains uncertain. TrustOps mitigates this uncertainty by fostering verifiable evidence of execution. Runtime observation tools, such as tracing mechanisms, container inspection and log aggregation tools, can be instrumented to collect and authenticate evidence, such as what versions are deployed, which servers handle a specific customer request and where data or logs are transferred to. This would turn so far vague statements in privacy policies\footnote{\enquote{Other optional tools in MongoDB Atlas require customer query log data to transit through our US-based servers.} -- \url{https://www.mongodb.com/legal/privacy}} into records every customer can verify. However, such transparency may create potential conflicts between the provider's need to keep business secrets and the public's need to verify regulatory compliance. To resolve such conflicts, TrustOps advocates for a synthesis of ZKPs or selective disclosure protocols to ensure user-relevant verifiability, without compromising vendors' proprietary information. Striking this balance necessitates advancements in observability tooling and evidence-collection methodologies, as well as a collaborative effort between service providers and users and a commitment from service vendors to enhance transparency and verifiability. While this shows how openly developed software can strongly benefit from TrustOps principles in both development and delivery, we also see benefits in internal organization development processes.

\subsection{Internal and External Organizational Trust}

Particularly concerning software companies with closed-source products and platforms, trustworthiness in the software ecosystem encompasses varying degrees of transparency. As technology advances, it becomes imperative for such entities to explore mechanisms ensuring verifiable trustworthiness, impacting reputation, lawfulness, and potential business development~\cite{ting2021trust}. The establishment of trust may occur both internally and externally, with organizational values and business decisions influencing the boundaries between these domains.

Internally, companies can strive for attestable quality of artifacts and accountability across teams and departments, incorporating elements like verifiable test pipelines, authenticated builds and releases, authorization, or monitoring infrastructure, during development, integration, or deployment. Trustworthiness requirements should be elicited before adopting such TrustOps practices, following a concrete trust and thread modelling approach. Moreover, software companies assess performance using diverse metrics, including pull request and code review statistics, or deployment and quality assessments, serving as potential operational evidence to be transformed into attested trust assurances. Automating attestation of these metrics fastens and strengthens trust in the delivered software and its building teams. However, this requires a judicious approach to prevent compromising other business aspects, for instance, in the spirit of site reliability engineering, with trust budgets instead of error budgets.

Externally, trust in software organizations is shaped by users and stakeholders, influenced by numerous factors, variables, and participants within the software ecosystem \cite{ting2021trust}. Nevertheless, a shift towards automated and verifiable claims holds the promise of translating abstract notions of trust into concrete data points. Verifiability of software and library versions, certified compliance with standards, and other recurring auditability tasks leading to public and organizational assessments are potential planes of automation and verifiability~\cite{fitzgerald2017continuous,myrbakken2017devsecops,enoiu2023veridevops}. Enabling the publication and verification of authenticated evidence generated during the TrustOps phases has the potential to increase organizational trustworthiness to internal and external stakeholders. Hence, to foster holistic trust in the services delivered, software companies can integrate TrustOps principles and follow the proposed approaches, but stakeholders may also have the agency to demand assurances regarding the incorporation of these practices, ensuring a balance between companies' business interests and everyone's privacy, security, and trust requirements.

%% file: sections/05_trustops_phases.tex
\begin{figure}
    \centering
    \includegraphics[width=0.9\textwidth]{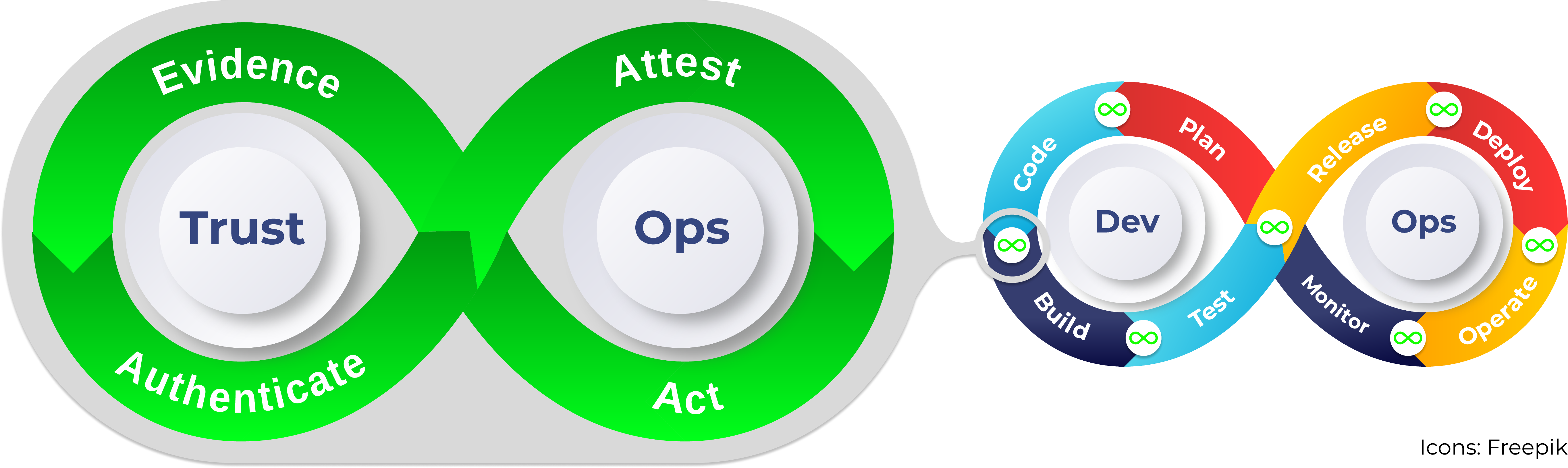}
    \caption{Overview of the idealized TrustOps life cycle, integrated within DevOps phases.}
    \label{fig:trustOps}
\end{figure}

% !TeX root = ../main.tex
\newcommand{\phase}[1]{\paragraph{\textbf{#1:}}}
TrustOps aims to enrich the software life cycle by continuously applying the evidence-based life cycle (\cref{ch:main}) throughout the phases of DevOps.
Overall, the aim is to accumulate and combine evidence from prior phases to achieve complex but automated attestation and authorization processes, thus enabling the possibilities sketched in \cref{ch:senarios}.
Similarly to DevOps itself, TrustOps can be applied as needed and can use as many or as few layers of trust-enhancing technologies as required. 
In the following, we present each of the DevOps Phases~\cite{bass2015devops} and discuss how TrustOps could be applied, exemplified in \Cref{fig:trustOps}.

\phase{Plan}
During the planning phase, features and changes are typically put into the software development backlog. 
These proposals are considered evidence in TrustOps and can be captured.
The aim is to determine how a change to a software artifact was introduced, starting with the issue or ticket that caused it.
Moreover, any policies related to the approval or change management process may be evident as well.

Once evidence is collected, it can be enriched to create authentic planning evidence, for example, by signing the issue or ticket, or providing immutable logs of the decision-making process. This allows the remainder of the development process to be linked to the original issue or ticket and, thus, the cause of a change, even if an issue or ticket is closed or deleted later.

Lastly, authenticated evidence should be attested to be used in conjunction with other planning tools or to authorize or trigger the execution of planned changes, e.g., by checking if a user story meets project standards, is signed by at least one senior developer, or was discussed in a sprint planning meeting.

\phase{Code}
Coding is the central part of any software project and, thus, also the central source of evidence in TrustOps.
At the core of this phase, code may be linked, in a verifiable way, to the developers that introduced the changes.

Moreover, considering the previous phase, we can also link code changes to planned changes, i.e., issues. 
Depending on the scenario, this may also include the collection of evidence about the development environment, usage of tools (e.g., linters, unit tests), and even the reliance on AI-based code assistants that could introduce intended or unintended changes.
Some software tools can already exploit prior attested planning evidence, for instance, by linking and referencing the issues in the commit messages and by authenticating the assigned developers. 
Similarly, development environments could be augmented to provide evidence of the environment, tools used, and tests run.

This authenticated evidence can then be used to create succinct proofs of correctness, e.g., verifying that a change was linted, tested, corresponds to a planned change, and is signed by the developer. Finally, successfully attesting to these proofs may trigger further phases in the development life cycle.

\phase{Build}
During the build phase, we follow the automated DevOps principles, which typically rely on CI/CD pipelines for building releasable software artifacts. 
Evidence represents the input to the build process, including the code and the configuration of the build system.
The build process may incorporate or connect the evidence from both the code and planning phases. This ensures that these typically self-contained artifacts remain authentic and attestable, in relation to the evidence from preceding phases and the TrustOps processes as a whole.

One way to produce authenticated builds in this phase would be to run the process inside TEEs, allowing later attestation of non-tampered builds.
However, other mechanisms could be similarly valid, depending on the needs of the specific software.
Additional attestations could combine and verify the collected evidence of all prior phases to ensure that a build fully complies with standards before pushing the build artifact to the test or release phases.

\phase{Test}
Some contexts require thoroughly tested software, and the record of these tests should be present when reviewing any released version.
Here, similar steps to the build process can be taken, e.g., recording evidence of the test environment and the tests that were actually run, or, if possible, even running tests in TEEs to attest to the test reports and ensure test results were not manipulated.

The aim is, thus, very similar to the build phase, to ensure that a tested build is backed by evidence of the prior phases and that the tests themselves are also backed by authenticated evidence.

\phase{Release}
Once an artifact is built and tested, it can be released.
Here, one of TrustOps' main goals can be seen, as now a released artifact can be backed by a chain of evidence that can be used to verify its authenticity and correctness, ensuring that the release is of claimed quality. 
Privacy-preserving technologies can also be used, e.g., by utilizing ZKPs, the adherence to specific project polices for testing and reviewing could be proven to the public, without revealing where tests were performed or which person reviewed changes, thus, removing the need to share confidential or private evidence without losing the ability to act on it.

Especially in fully open and public development scenarios, this provides the opportunity to attach evidence to publicly verifiable proofs. These proofs can be checked for compliance, beyond simple hash-based integrity, before they are used as dependencies elsewhere, e.g., to make supply chain attacks more costly.

\phase{Deploy}
During the deployment phase, artifacts are typically deployed for use. 
Here, the extent of TrustOps involvement strongly depends on the type of application.
In most cases, we assume that both the build artifacts as well as the deployment code are part of the same repository and, thus, are backed by clear lineage on how changes to the deployment were made. 
However, during this phase, some additional environment information may be needed to turn a deployment template into a concrete and verifiable deployment. 

At the minimum, however, we could collect who authorized a deployment and collect a record of what artifact is deployed, including the verification of its correctness and authenticity that was built up in the prior phases.
Furthermore, this phase could provide evidence of the deployment environment, e.g., outputs of reproducible deployment mechanisms such as NIX or Docker images and the configuration of the deployment environment.
During deployment, the management of the identity of the environment and operators is also a critical component that may require attestation. 

\phase{Operate}
During the operating phase, it is critical to enable reasonable observability to collect authentic evidence.
This may include the verification of execution environments and the identification of users. 
Moreover, this phase will likely produce the highest amount of raw evidence, typically in the form of logs, metrics, or traces.
Hence, the evidence collected in this phase is likely to be the most diverse and the most sensitive. 
Thus, the use of privacy-preserving mechanisms is a key enabler in managing the amount and sensitivity of operational raw evidence. However, developers and operators should make sure that only the necessary evidence is collected, authenticated, and attested.
The evidence collected may also be linked to all prior evidence that led to the current state of the software system.
Therefore, this phase may require a new category of tooling to support the efficient collection of runtime attestable evidence that goes beyond existing observability solutions.

\phase{Monitor}
In this last phase, the collected evidence can be used to link and provide verifiability feedback to users or operators, e.g., giving them the means to automatically attest if a software deployment is the same as the one they expect, operating in the way they expect, thus, providing the necessary assurances to trust the running software and its operators.

The evidence collected in this and prior phases can be used to feed the cycle again, as already authenticated input for the planning phase, e.g., utilizing the built chain of evidence as clear information on what interaction caused a fault in the running software that should be fixed.

%% file: sections/06_research_challenges.tex
This section highlights the main foreseeable challenges of adopting TrustOps, in three categories: evidence management, TrustOps integration and usability.

Evidence collected for TrustOps may contain highly sensitive information or be used for other purposes than TrustOps.
Hence, we must consider how to store, manage and use evidence in a privacy-preserving manner while minimizing the overhead of handling it. 
Different types and life cycle stages of evidence make proposing a standard for evidence management challenging. Different evidence generation and management tools can also cause interoperability issues. 
Consequently, further work must establish best practices for which evidence to collect and how to manage it.

For TrustOps to be fully integrated into existing practices, it must: a) coexist with applications that do not use or adhere to the proposed method or require particular strategies, and b) establish recommended tools, practices, standards, and guidelines.
This necessitates the development of new tools or extensions within existing development or operational environments to effectively support TrustOps across the phases of the software life cycle.

The usability of underlying technologies like ZKPs and TEEs must improve to support the use of TrustOps. This includes improvements tailored for developers and operators to streamline integration processes.
Moreover, a thorough threat model analysis and accompanying recommendations are essential for practitioners. These insights should help guide decisions regarding the onboarding of TrustOps. In addition, possible users should be educated regarding the capabilities and shortcomings of  verification tools.

%% file: sections/07_conclusion.tex
In this paper, we presented TrustOps, an extension of DevOps that integrates evidence verification and identity management into the software development process. TrustOps aims to build and maintain trust by making software actions transparent, understandable, and verifiable. It adopts an evidence-based approach, focusing on generating, enhancing, and providing authenticated, attested, and actionable evidence at each step of the software life cycle.

We discussed application scenarios demonstrating how TrustOps principles enhance security, transparency, and trust in various contexts, including open-source software, service ecosystems, and organizational development. Key research challenges include evidence management, integration and adoption, and usability and understanding. Through these concepts and challenges, we provide a roadmap for the software engineering, security, and privacy communities, as well as the industry, to advance the TrustOps approach.

Despite the challenges, the potential benefits of TrustOps in enhancing trust, transparency, and accountability make it a promising direction for future research. We already see several emerging projects that offer solutions to cover some aspects of TrustOps\footnote{A continuously updating collection of these resources is available here:\\  \url{https://github.com/trustops/awesome-trustops}}. By embracing verifiable development practices and the principles described, TrustOps could improve software development and significantly enhance trust in software systems, across domains and applications.